\documentclass[%
 reprint,
superscriptaddress,
 amsmath,amssymb,
 aps,
]{revtex4-2}

\usepackage{graphicx}
\usepackage{dcolumn}
\usepackage{bm}
\usepackage[capitalize]{cleveref}
\usepackage{chemformula}
\usepackage[version=3]{mhchem} 
\usepackage{siunitx}

\begin{document}

\preprint{APS/123-QED}

\title{Dynamic strain modulation of a nanowire quantum dot compatible with a thin-film lithium niobate photonic platform}

\author{Thomas Descamps}
\email{descamps@kth.se}
 \affiliation{Department of Applied Physics, KTH Royal Institute of Technology, Roslagstullsbacken 21, 10691 Stockholm, Sweden}
\author{Tanguy Schetelat}
 \affiliation{Department of Applied Physics, KTH Royal Institute of Technology, Roslagstullsbacken 21, 10691 Stockholm, Sweden}
\author{Jun Gao}
 \affiliation{Department of Applied Physics, KTH Royal Institute of Technology, Roslagstullsbacken 21, 10691 Stockholm, Sweden}
 \author{Philip J. Poole}
 \affiliation{National Research Council of Canada, Ottawa, Ontario K1A 0R6, Canada}
 \author{Dan Dalacu}
\affiliation{National Research Council of Canada, Ottawa, Ontario K1A 0R6, Canada}
\author{Ali W. Elshaari}
 \affiliation{Department of Applied Physics, KTH Royal Institute of Technology, Roslagstullsbacken 21, 10691 Stockholm, Sweden}
\author{Val Zwiller}
\email{zwiller@kth.se}
\affiliation{Department of Applied Physics, KTH Royal Institute of Technology, Roslagstullsbacken 21, 10691 Stockholm, Sweden}
\affiliation{Single Quantum BV, Delft, The Netherlands}

\date{\today}

\begin{abstract}
The integration of on-demand single photon sources in photonic circuits is a major prerequisite for on-chip quantum applications. Among the various high-quality sources, nanowire quantum dots can be efficiently coupled to optical waveguides because of their preferred emission direction along their growth direction. However, local tuning of the emission properties remains challenging. In this work, we transfer a nanowire quantum dot on a bulk lithium niobate substrate and show that its emission can be dynamically tuned by acousto-optical coupling with surface acoustic waves. The purity of the single photon source is preserved during the strain modulation. We further demonstrate that the transduction is maintained even with a \ch{SiO_2} encapsulation layer deposited on top of the nanowire acting as the cladding of a photonic circuit. Based on these experimental findings and numerical simulations, we introduce a device architecture consisting of a nanowire quantum dot efficiently coupled to a thin film lithium niobate rib waveguide and strain-tunable by surface acoustic waves. 

\end{abstract}

\maketitle



Quantum computation with on-chip photonic qubits has emerged as an important research area in quantum technologies \cite{OBrien2007, Zhong2020, Elshaari2020a, Pelucchi2021}. In addition to the manipulation and detection of individual photons, achieving deterministic integration of high-purity, on-demand single photon sources is essential for the scalability and complexity of quantum photonic circuits \cite{Bounouar2020, Rodt2021, Moody2022}. Although more technologically challenging than monolithic approaches \cite{Dietrich2016a, Lenzini2018, Li2023}, the heterogeneous integration of high-quality quantum emitters offers the possibility of producing more efficient devices and incorporating additional functionalities that would be difficult to achieve in single material systems. Towards this direction, III/V semiconductor quantum dots (QDs) \cite{Chanana2022, Aghaeimeibodi2018, Katsumi2021, Elshaari2017, Mnaymneh2020}, and defects in crystals \cite{Wan2020, Guo2021} or in 2D materials \cite{White2019, Peyskens2019, Liu2023} have been successfully transferred to Si, SiN, AlN or \ce{LiNbO_3} while maintaining good performances as single photon sources. In addition to their individual properties, multiple sources integrated on the same chip must be spectrally identical if indistinguishable photons are to be generated for linear quantum operations \cite{Rodt2021}. Using a monolithic approach, two-photon interference between photons generated by two self-assembled QDs located in two different waveguides has recently been shown \cite{Dusanowski2023}. Although surface scanning was successfully used to identify two QDs with naturally good spectral overlap, this method is unlikely to be efficient for QDs transferred to a host substrate. Indeed, even if the sources are carefully preselected on their original substrate, their spectral properties may change after integration due to a different charge and strain environment. Consequently, including an external tuning scheme for each source is essential to bring them into spectral resonance. Besides, finding naturally identical emitters is not a scalable approach, so having a tuning knob is desirable to relax the preselection step. Multiple tuning schemes have been explored based on temperature \cite{Faraon2009}, magnetic field \cite{Bayer2002}, electrical field via the quantum confined Stark effect \cite{Schnauber2021}, nanomechanical systems \cite{Midolo2018}, strain field with piezoelectric materials \cite{Zhang2016} and surface acoustic waves (SAWs). 
The latter are generated from a piezoelectric material by an interdigital transducer (IDT) and can interact with a large variety of quantum systems. Coupling to superconducting qubits \cite{Dumur2021, Bienfait2019}, coherent electron transport between gate-defined QDs \cite{Jadot2021} as well as coherent acoustic control of optically active QDs \cite{Imany2022, Weiß2021} have been demonstrated. In the latter case, the piezoelectric potential generated by the SAW can be used to transport photo-generated carriers to the QD \cite{Hernandez-Minguez2012}, while the QD bandgap can also be modulated by the in-plane strain field of the SAW \cite{Villa2017, Tiwari2020, Lazic2019, Buhler2022}.

\begin{figure*}[ht]
\includegraphics[width=0.9\textwidth]{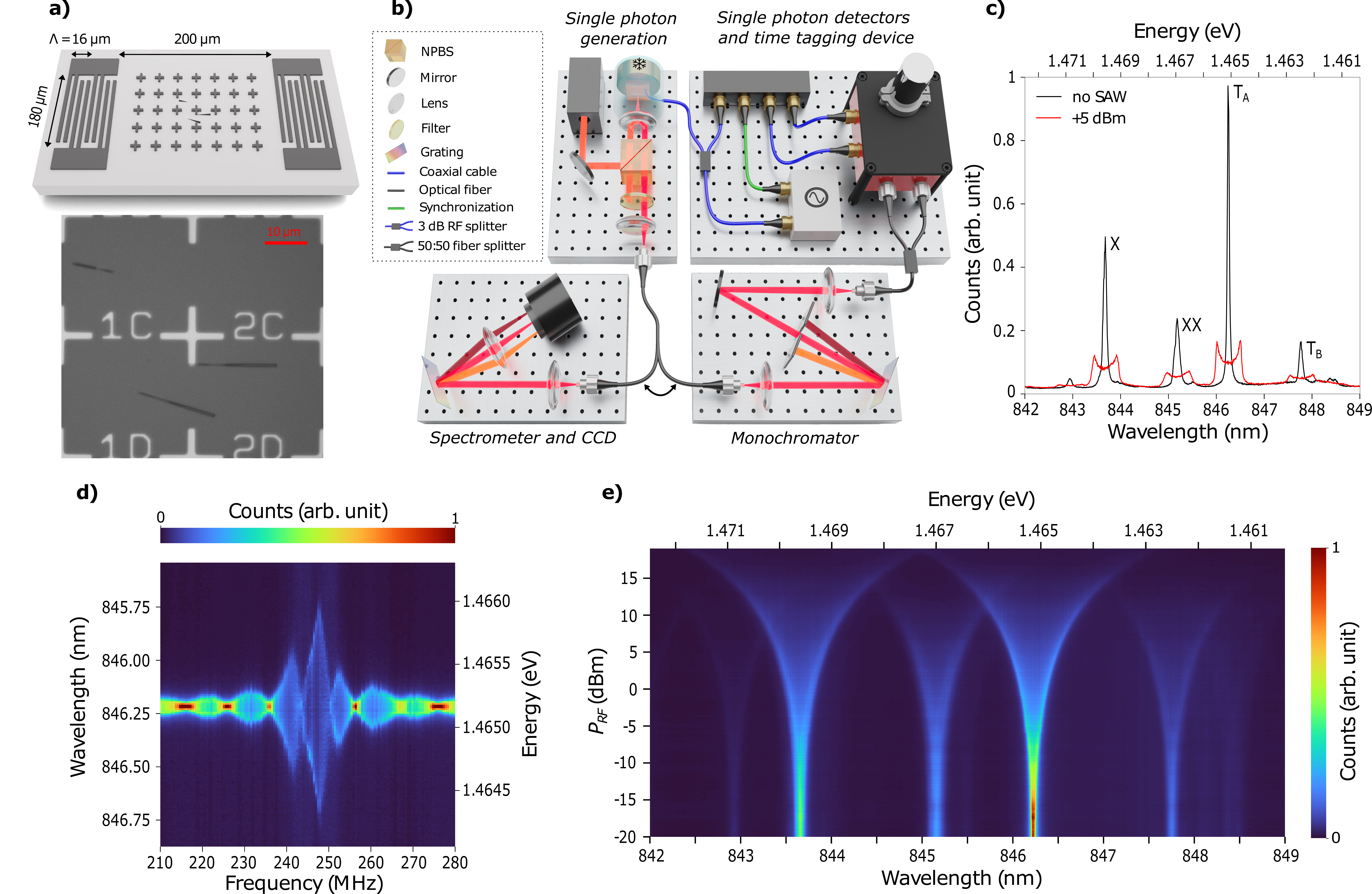}
\caption{\label{fig:CW_measurement_woSIO2}a) Sketch of the surface acoustic wave delay line with an optical microscope image of individually transferred nanowires. b) Optical setup for spectroscopy or time-resolved measurement (NPBS: non-polarizing beam splitter). c) Emission peaks of the nanowire QD without (black) and with (red) the SAW-induced modulation ($P_{RF} = \SI{5}{dBm}$ at $\SI{247.5}{MHz}$). d) Optical modulation of line $T_A$ around the fundamental resonance frequency of the IDT ($P_{RF} = \SI{10}{dBm}$). e) Optical modulation as a function of $P_{RF}$ (SAW generation at $\SI{247.5}{MHz}$). For all measurements, the QD was continuously excited with a He-Ne laser at $\SI{150}{nW}$.}
\end{figure*}

Among the various transferable emitters or defects, InAsP QDs embedded in an InP nanowire (NW) with a taper-shaped InP shell stands out as a good candidate \cite{Chang2023, Gao2023}. Bright \cite{Reimer2012}, high-purity \cite{Dalacu2012}, near transform-limited \cite{Laferriere2023} and indistinguishable single photons as well as entangled photons pairs \cite{Versteegh2014,Jons2017} have been preferentially emitted along the long axis of the NW with a Gaussian transverse mode. This emission profile is advantageous for efficient coupling to single mode waveguides and fibers. Site-controlled growth with high yield \cite{Laferriere2022} is also particularly convenient for deterministic integration on photonic chips using pick-and-place techniques. On-chip routing, filtering, tuning, multiplexing and detection of emitted single photons has been demonstrated on a SiN platform \cite{Elshaari2017, Mnaymneh2020, Elshaari2018, Zadeh2016, Gourgues2019}.
In this work, we demonstrate the modulation of a single tapered nanowire QD with surface acoustic waves. The NW was transferred to a bulk \ce{LiNbO_3} substrate and bonded to the substrate by Van der Waals interactions. \ce{LiNbO_3} offers a large electromechanical coupling coefficient ($k^{2} \approx \SI{5}{\percent}$) making it particularly suited to modulate the emission of QDs at moderate RF power \cite{Nysten2017}. We show that the emission wavelength of the quantum emitter can be tuned by more than $\SI{1}{nm}$ at the SAW resonance frequency while the single photon purity is preserved. We then explore the influence of a \ce{SiO_2} encapsulation layer on the strain tuning performance. On the one side, this layer rigidly anchors the NW to the substrate, making it easier to manufacture a photonic chip if the NW is first transferred onto an unprocessed substrate. On the other side, once the NW is placed on the photonic waveguide, the \ce{SiO_2} layer acts as a cladding layer. Based on these findings, we propose a device architecture where the NW is adiabatically coupled to a rib waveguide made from a \ce{LiNbO_3} thin film and compatible with SAW-induced strain modulation. This fully integrated strain-tunable source will open the way for further on-chip manipulation and detection of photonic qubits.

\section{Experiments}

The chip consists of two IDTs forming a $\SI{200}{\micro m}$ delay line patterned on a 128$^{\circ}$ Y-X cut \ce{LiNbO_3} substrate by optical lithography and etching of a $\SI{60}{nm}$-thick chromium layer (\cref{fig:CW_measurement_woSIO2}(a)). The IDTs have a double-electrode structure to avoid internal reflections of the SAW \cite{Ralib2015}. The electrodes are $\SI{2}{\micro m}$ wide, resulting in a spatial period $\Lambda=\SI{16}{\micro m}$ repeated 25 times per IDT. This configuration gives two SAW excitation frequencies, one fundamental resonance at $f_{1} = c_{s}/\Lambda = \SI{249}{MHz}$ and a third harmonic at $f_{3} = 3c_{s}/\Lambda = \SI{748}{MHz}$ where $c_{s}$ is the speed of sound in the substrate ($c_{\ce{LiNbO_3}} = \SI{3990}{m s^{-1}}$). The site-controlled InP NW embedding the InAsP QD was individually picked up with a micro-manipulator and transferred \cite{Mnaymneh2020} onto the \ce{LiNbO_3} chip within the delay line. The long axis of the NW is approximately parallel to the direction of propagation of the acoustic wave. In the following measurements, only the fundamental resonance is investigated.
The sample was investigated at $\SI{1.8}{K}$ in a dry cryostat designed for confocal micro-photoluminescence (PL) measurements and equipped with high frequency cables (\cref{fig:CW_measurement_woSIO2}(b)). A continuous wave He-Ne laser was focused with a microscope objective to excite the QD above band. The collected PL was dispersed by a $\SI{750}{cm}$ focal length spectrometer and detected by a liquid nitrogen-cooled charge-coupled device (CCD) camera. An analog signal generator was used to apply a sinusoidal radio frequency (RF) signal of adjustable power $P_{RF}$ to one IDT of the delay line while the other was floating.
\cref{fig:CW_measurement_woSIO2}(c) shows the PL spectrum of the QD without applying SAWs. From higher to lower energy, the four peaks are identified as the exciton $X$, the biexciton $XX$ and two trions $T_{A}$ and $T_{B}$ based on power-dependent measurements (see Supporting Information S1) \cite{Weiß2014} and previous studies \cite{Versteegh2014}. 

When a RF signal is applied to the IDT, the radiated SAW couples to the QD through its strain field. Since the QD size is smaller by approximately three orders of magnitude than the wavelength $\Lambda$ of the SAW, the strain field experienced by the QD can be considered uniformly distributed. The sinusoidal modulation of the strain field causes a modulation of the bandgap of the QD at the same frequency, resulting in the oscillation of the spectral lines around their unstrained energies. The time average of this oscillation gives a broad spectral line with peaks at the edges, as plotted in \cref{fig:CW_measurement_woSIO2}(c) for a $P_{RF}=\SI{5}{dBm}$ sinusoidal signal at $\SI{247.5}{MHz}$. For a given RF power, the tuning magnitude is largest when the SAW is generated close to the resonance frequency $f_{1}$ of the IDT, as shown in \cref{fig:CW_measurement_woSIO2}(d) for the $T_A$ line. The main resonance peak is centered around $\SI{247.5}{MHz}$, which corresponds to an effective speed of sound in \ce{LiNbO_3} of $c_{\ce{LiNbO_3}, \text{eff}} = \SI{3960}{ms^{-1}}$. This main resonance spanning $\SI{20}{MHz}$ is split into three lobes separated by $\delta f = \SI{6.6}{MHz}$ since the delay line behaves as an acoustic cavity. The effective length of the cavity is hence $c_{\ce{LiNbO_3}, \text{eff}}/2\delta f \simeq \SI{300}{\micro m}$ which is slightly larger than the physical length of the delay line due to the penetration of the SAW inside the two IDTs. A similar response is obtained for all the other lines of the QD.
At the SAW resonance frequency, detuning becomes observable for all the peaks around $P_{RF}=\SI{-15}{dBm}$ and can reach up to $\SI{1.5}{nm}$ at $\SI{+15}{dBm}$ (\cref{fig:CW_measurement_woSIO2}(e)). The broadening remains symmetric around the unstrained emission, indicating that high RF powers can be applied without inducing local heating of the QD. A good mechanical contact between the NW and the substrate is also maintained as the detuning does not drop even at high RF powers. However, the integrated lineshape decrease for all peaks slightly above 5 dBm (see Supporting Information S2). The next experiments were carried out below this threshold.

\begin{figure}
\includegraphics[width=0.9\linewidth]{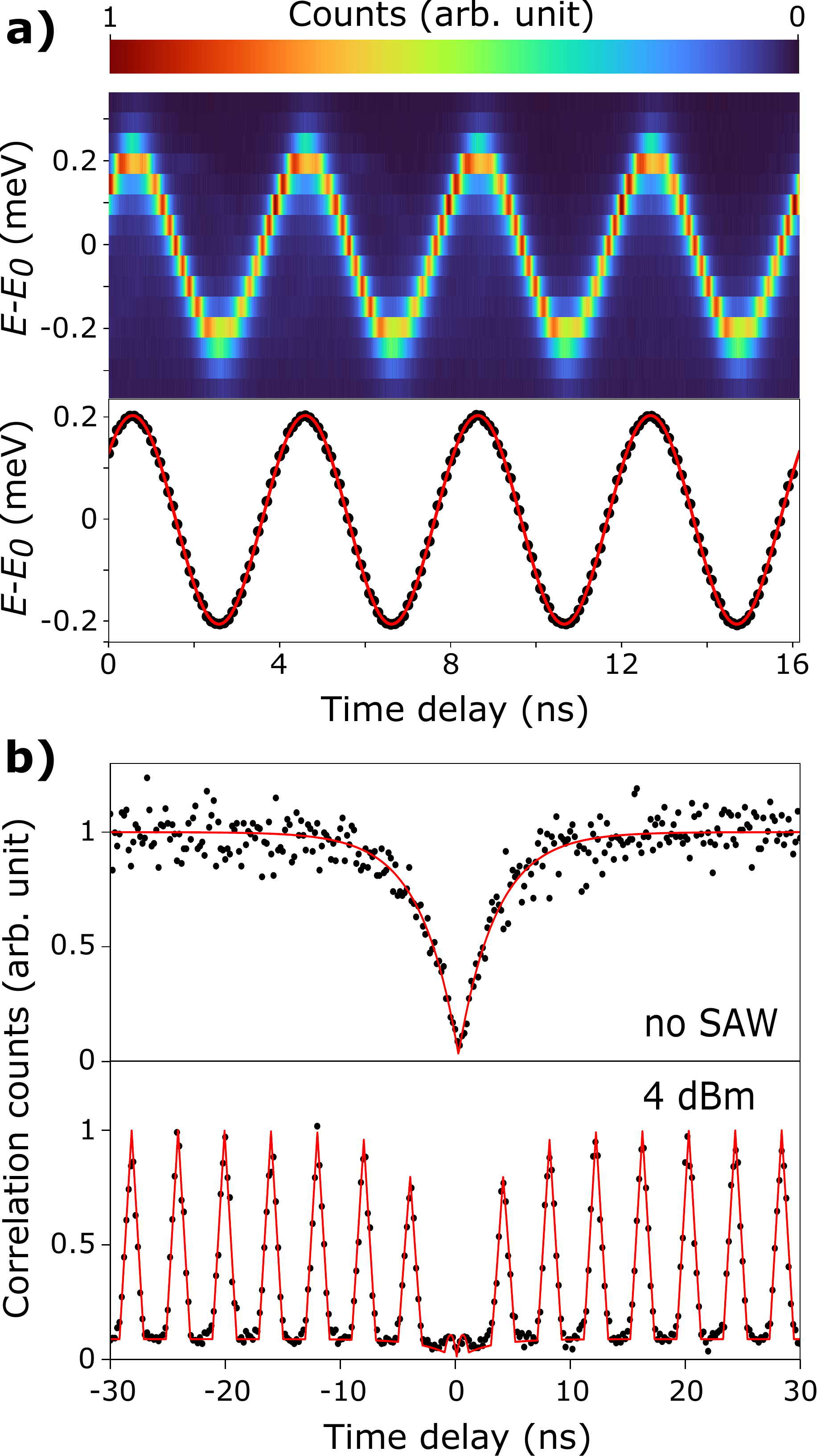}
\caption{\label{fig:TimeRes_measurement_WoSiO2} a) PL spectra of the $T_A$ line measured over 4 acoustic periods (top). The modulation with respect to the unstrained energy $E_0$ was extracted and fitted by a sine function (bottom) with a period of 4.038 ns. b) Second order correlation function of the $T_A$ line without (top) and with (bottom) SAW interaction. In the latter case, the monochromator filtered at the energy corresponding to the maximum of the sine wave shown in a). The fitting function (red lines) are detailed in the main text. For all the measurements, the QD was continuously excited with a HeNe laser at $\SI{150}{nW}$.}
\end{figure}

Time-resolved measurements were then performed to observe the strain tuning over one acoustic cycle. As drawn in \cref{fig:CW_measurement_woSIO2}(b), the PL signal was filtered by a monochromator ($\SI{0.1}{nm}$ bandwidth) around the $T_A$ emission line, detected by superconducting nanowire single photon detectors (time jitter around $\SI{20}{ps}$) and counted by a time tagging device (time jitter below $\SI{10}{ps}$) phase-locked to the signal generator. The sample was continuously excited with a He-Ne laser. A balanced power splitter was inserted at the output of the signal generator set at $\SI{247.5}{MHz}$ with output power $\SI{8}{dBm}$. Half of the power triggered the counting module while the other half drove the IDT.
Over an acoustic cycle, pulses in the count rate can be observed as soon as the $T_A$ emission wavelength falls within the filtering bandwidth of the monochromator. By sweeping the filtering window across the strain-tuned emission line, the time-dependent modulation can be reconstructed as shown in \cref{fig:TimeRes_measurement_WoSiO2}(a). This modulation is sinusoidal with a fitted frequency of $\SI{247.64}{MHz}$, which closely matches the RF drive frequency. The absence of distortions in the sine wave shows that the tuning is only induced by the deformation potential, without contribution from the piezoelectric field via the quantum confined Stark effect \cite{Weiß2014}.
The purity of the single photon source was then assessed by measuring the second-order correlation function of the $T_A$ line in a Hanbury Brown-Twiss measurement (\cref{fig:TimeRes_measurement_WoSiO2}(b)). The QD was continuously excited non-resonantly with a He-Ne laser. Without strain-tuning, $g^{(2)}(0) = 0.058$ is measured experimentally, whereas a value of $0.0313 \pm 0.019$ is found by fitting the data to a biexponential decay function, hence demonstrating the high purity of the emitter. A SAW was then generated at $P_{RF}=\SI{4}{dBm}$ while the PL was filtered at $\SI{846.58}{nm}$, which corresponds to the wavelength where the modulation is maximum. Pulses were measured with a period of $\SI{4.039}{ns}$ and the peak at zero time delay was strongly suppressed. The experimental data were fit based on the detection probability model introduced by \citet{Gell2008} in which $g^{(2)}(\tau)=P_e(\tau)P_d(\tau)$, where $P_e(\tau)$ is the probability of the QD emitting a pair of photons with a temporal separation of $\tau$ (biexponential decay function), and $P_d(\tau)$ is the pair-detection probability (convolution of two trains of top hat functions). The experimental point at zero delay $g^{(2)}(0) = 0.042$ and the fitting parameter $0.0114 \pm 0.026$ are very similar to the values obtained without modulation. This measurement under continuous wave excitation demonstrates the pulsed generation of high purity single photons over the tuning range induced by the strain modulation. A similar measurement was carried out when the monochromator filtered at $\SI{846.2}{nm}$ (Supporting Information S3), and shows that the period of the pulses is divided by two since the modulated emission falls within the monochromator band pass twice over one acoustic cycle.

The ability to generate indistinguishable photons is also an important requirement for single photon sources, and is typically measured in a Hong-Ou-Mandel two-photon interference experiment. Visibility as high as $\SI{83}{\percent}$ has been demonstrated with the same nanowire QD system as used in this study \cite{Reimer2016}. More recently, near Fourier transform-limited photons have been shown by carefully optimizing the growth conditions \cite{Laferriere2023}. For those two experiments, the nanowire QDs were standing up on the original surface and excited above-band. Here, the nanowire QD studied was grown under the same optimized conditions, except for a $\SI{15}{\celsius}$ lower growth temperature of the core. After transferring the nanowire on the \ce{LiNbO_3} substrate, the linewidth might broaden due to the proximity of the nanowire QD to a surface with a higher density of charges, degrading the degree of indistinguishability.
Although not performed in this study, two-photon interference using the photons emitted by the strain-tuned nanowire could be carried out with the same spectral filtering as that implemented for the previously detailed time-resolved experiments. It is also compatible with resonant excitation schemes in order to reach high visibility by detuning the laser to the same wavelength as the filtering window \cite{Villa2017} and by rejecting it at the monochromator output before the interferometer. Note that to avoid a potential change in polarization with the strain field, only one photon should be collected per acoustic cycle.

\begin{figure}[ht]
\includegraphics[width=0.9\linewidth]{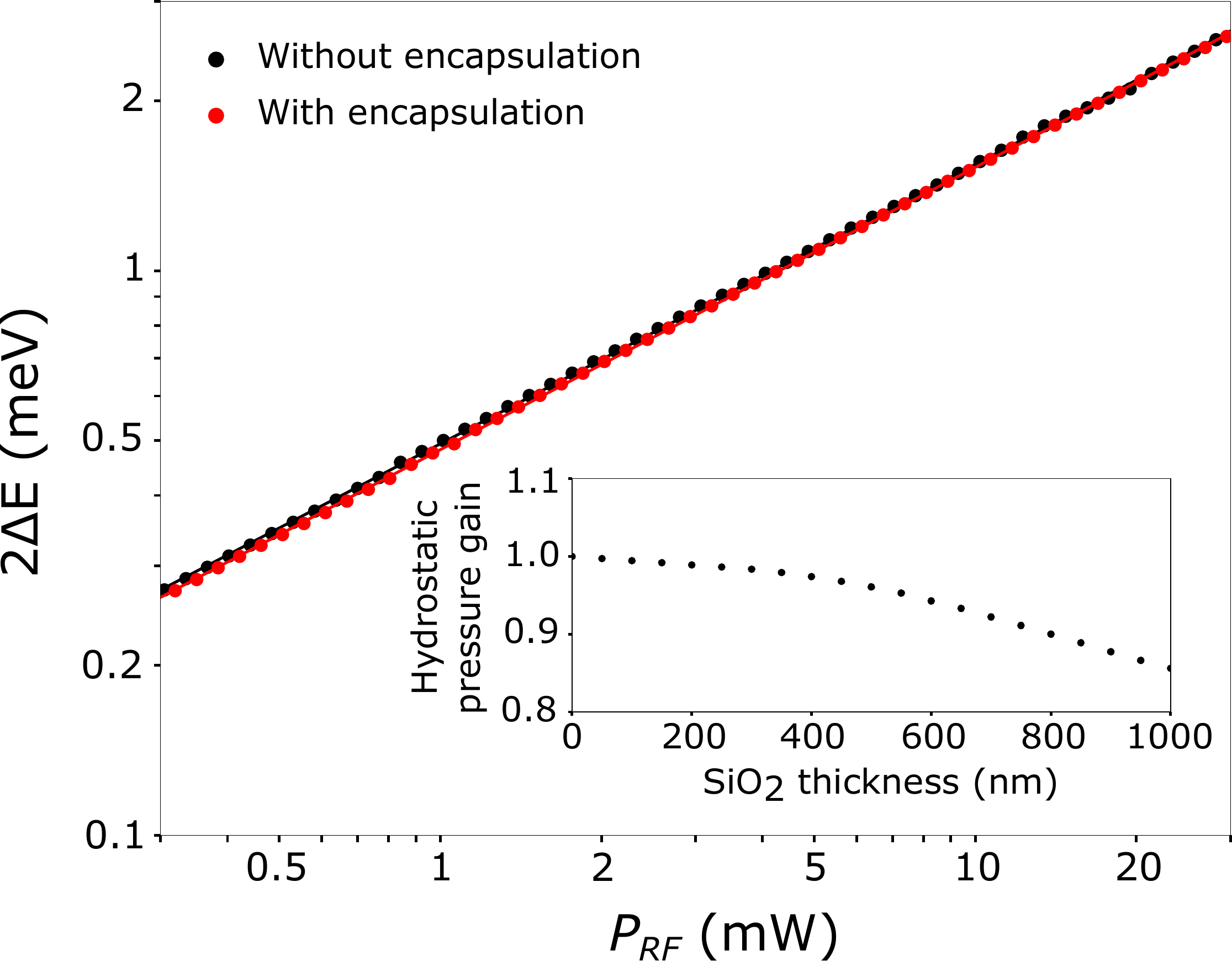}
\caption{\label{fig:influenceSiO2} Strain-induced energy splitting of the $T_A$ line as a function of $P_{RF}$. The black and red points were taken before and after the deposition of \ce{SiO_{2}} ($\SI{320}{nm}$) on top of the NW, respectively. The solid lines are linear fits. In both cases, the excitation power was $\SI{150}{nW}$. The inset shows the evolution of the simulated hydrostatic pressure in the InP layer after the deposition of a \ce{SiO_2} layer on top of the InP/\ce{LiNbO_3} stack.}
\end{figure}

Next, the tuning of the QD emission was investigated after depositing a $\SI{320}{nm}$-thick layer of \ce{SiO_2} by PECVD on the chip. The oxide on top of the IDTs was locally removed to have the same resonance frequency and the same electrical to acoustic transduction as previously (see Supporting Information S4 for a comparison of the scattering parameters). This oxide layer bonds the NW on the substrate more rigidly than the Van der Waals forces which provides two benefits. On the one hand, it eases the fabrication of subsequent photonic structures, and on the other hand, it plays the role of a cladding layer when the NW is integrated with a lithium niobate on insulator (LNOI) waveguide. 
The impact of the \ce{SiO_2} layer on the modulation was quantified by measuring the energy broadening as a function of RF power, similarly to \cref{fig:CW_measurement_woSIO2}(c). The driving frequency chosen to maximize the acousto-optical coupling was still $\SI{247.5}{MHz}$. Independent of the SAW power, the PL lines were blueshifted by $\SI{1.92}{nm}$, indicating a static strain introduced by the \ce{SiO_2} layer onto the QD. Without the SAW, no significant broadening of the QD emission line could be resolved after encapsulation. The optomechanical modulation amplitude $2\Delta E$ was extracted from the data by fitting to a time-integrated oscillating Lorentzian emission line \cite{Nysten2020}. The optomechanical response $2\Delta E$ is plotted as a function of the driving RF power in logarithmic scale before and after \ce{SiO_2} encapsulation (\cref{fig:influenceSiO2}). In both cases, the strain-induced broadening follows the power law $2\Delta E \propto (P_{\text{RF}})^{\alpha}$, where $\alpha=0.5001 \pm 0.0003$ with \ce{SiO_2} and $\alpha=0.4951 \pm 0.0004$ without. These two coefficients are very close to the ideal value $\alpha = 0.5$ for a deformation potential coupling \cite{Pustiowski2015}, demonstrating that the observed broadening is essentially due to optomechanical coupling. For a given RF power within the range $\SI{-5}{dBm}$ to $\SI{15}{dBm}$, the encapsulation resulted on average in a broadening smaller by around $\SI{1.8}{\percent}$. 
The influence of the encapsulating layer thickness on the acousto-optical coupling was studied based on 2D FEM frequency domain simulations with COMSOL. The SAW was excited at the fundamental resonance of a bulk 128$^{\circ}$ Y-X cut \ce{LiNbO_3} substrate, and propagated along the X-axis towards a NW placed on top of the substrate. The NW is modeled as a linear elastic slab of InP with 200 nm thickness, corresponding to the diameter of the NW at the location of the quantum dot. A linear elastic \ce{SiO_2} layer of variable thickness was added on top of this slab. The SAW-induced hydrostatic pressure $p$ at the center of InP layer defined as $p=-E_Y \varepsilon$, where $\varepsilon$ and $E_Y$ are the trace of the strain tensor and the Young's modulus respectively, was computed at the center of the InP layer \cite{Weiß2018a}. The density of the PECVD oxide was set to $\SI{2200}{kg/m^{3}}$, its Poisson's ratio to 0.24 \cite{Carlotti1997, Doucet1995} and the Young's modulus was considered as a fitting parameter to match our model to the experimental data. A Young's modulus of $\SI{82}{GPa}$ was obtained for our film. The influence of the \ce{SiO_2} thickness on the hydrostatic pressure inside the \ce{InP} layer was then simulated and is shown in the inset of \cref{fig:influenceSiO2}. For thicker encapsulation layers, the hydrostatic pressure decreases in the \ce{InP} layer as the SAW is no longer localized at the top of the bulk \ce{LiNbO_3} but also propagates at the surface of the \ce{SiO_2} (see Supporting Information S5 for strain profiles).

\section{Discussion}

\begin{figure}
\includegraphics[width=\linewidth]{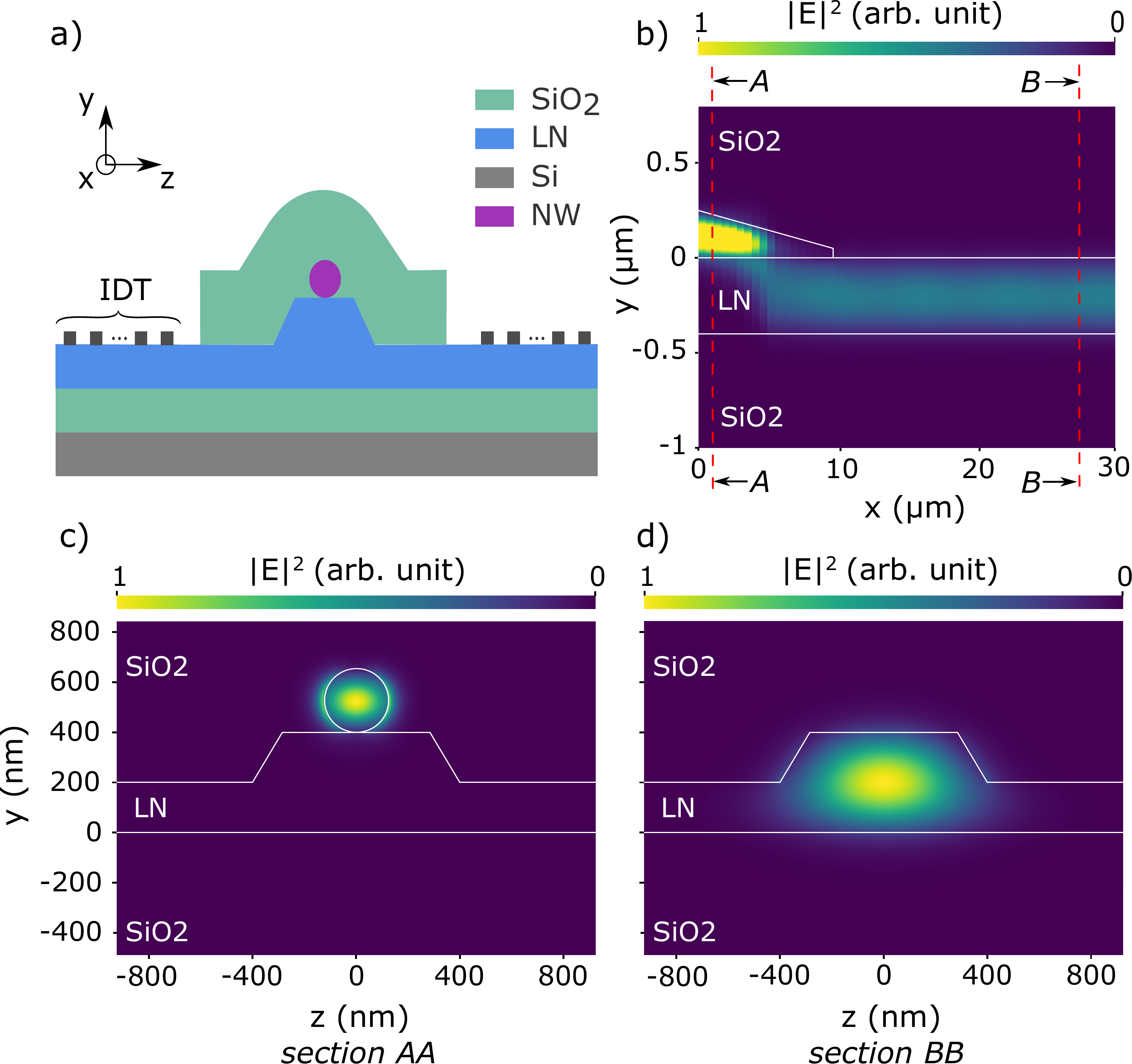}
\caption{\label{fig:modes} a) Sketch of the nanowire QD placed on top of a thin film \ce{LiNbO_3} rib waveguide with IDTs on each side for acousto-optical modulation. The \ce{SiO_2} layer acting as cladding and encapsulation layer for the nanowire is locally removed in the IDT region. b) Optical TE mode transfer between the tapered NW and the rib waveguide. c) Cross section AA showing the confined optical mode in the NW before the mode transfer region. d) Cross section BB representing the optical mode confined in the rib waveguide after the mode transfer region. The refractive indices at $\SI{850}{nm}$ are $n_{\ce{LiNbO_3}}$(e) = 2.17, $n_{\ce{LiNbO_3}}$(o) = 2.25, $n_{\ce{SiO_2}}$ = 1.45, and $n_{\ce{InP}}$ = 3.46. The axes correspond to that of a Y-cut LNOI.}
\end{figure}

Guiding the light emitted by the QD and modulated by SAWs would enable on-chip single photon manipulation and detection. Towards this direction, we propose a novel architecture shown in \cref{fig:modes}(a). The tapered NW is placed on top of an X-oriented rib waveguide made from a $\SI{400}{nm}$-thick Y-cut LNOI wafer. The NW is modeled as a $\SI{10}{\micro m}$-long truncated cone with base diameter $\SI{200}{nm}$ and top diameter $\SI{50}{nm}$. The rib waveguide has a height of $\SI{200}{nm}$, leaving a base of $\SI{200}{nm}$ to drive SAWs. The thick buried oxide ($\SI{3}{\micro m}$) efficiently confines the optical mode in the rib waveguide, while the NW is encapsulated by a $\SI{320}{nm}$ thick conformal \ce{SiO_2} layer. 
The tapered shape of the NW favors an adiabatic mode transfer of the TE and TM modes of the NW \cite{Mnaymneh2020} to the fundamental TE and TM modes of the waveguide, respectively (\cref{fig:modes}(b-d)). Finite-difference time-domain simulations (Lumerical) were conducted to calculate the coupling efficiency of the device assuming lossless materials. For a $\SI{800}{nm}$-wide rib waveguide, a coupling efficiency of $\SI{95.1}{\percent}$ for the TE modes and $\SI{98}{\percent}$ for the TM modes was obtained. This large waveguide geometry further facilitates the placement of the NW using a manipulator. The efficiency of other waveguide geometries can be found in Supporting Information S6. The IDTs excite SAWs along the Z-axis of the crystal which has the largest electromechanical coupling coefficient. Unlike in the experimental section, the NW is now perpendicular to the propagation direction of the SAW, but this change in orientation does not significantly affect the strain-induced coupling as shown in the Supporting Information S7. 

\begin{figure}
\includegraphics[width=0.9\linewidth]{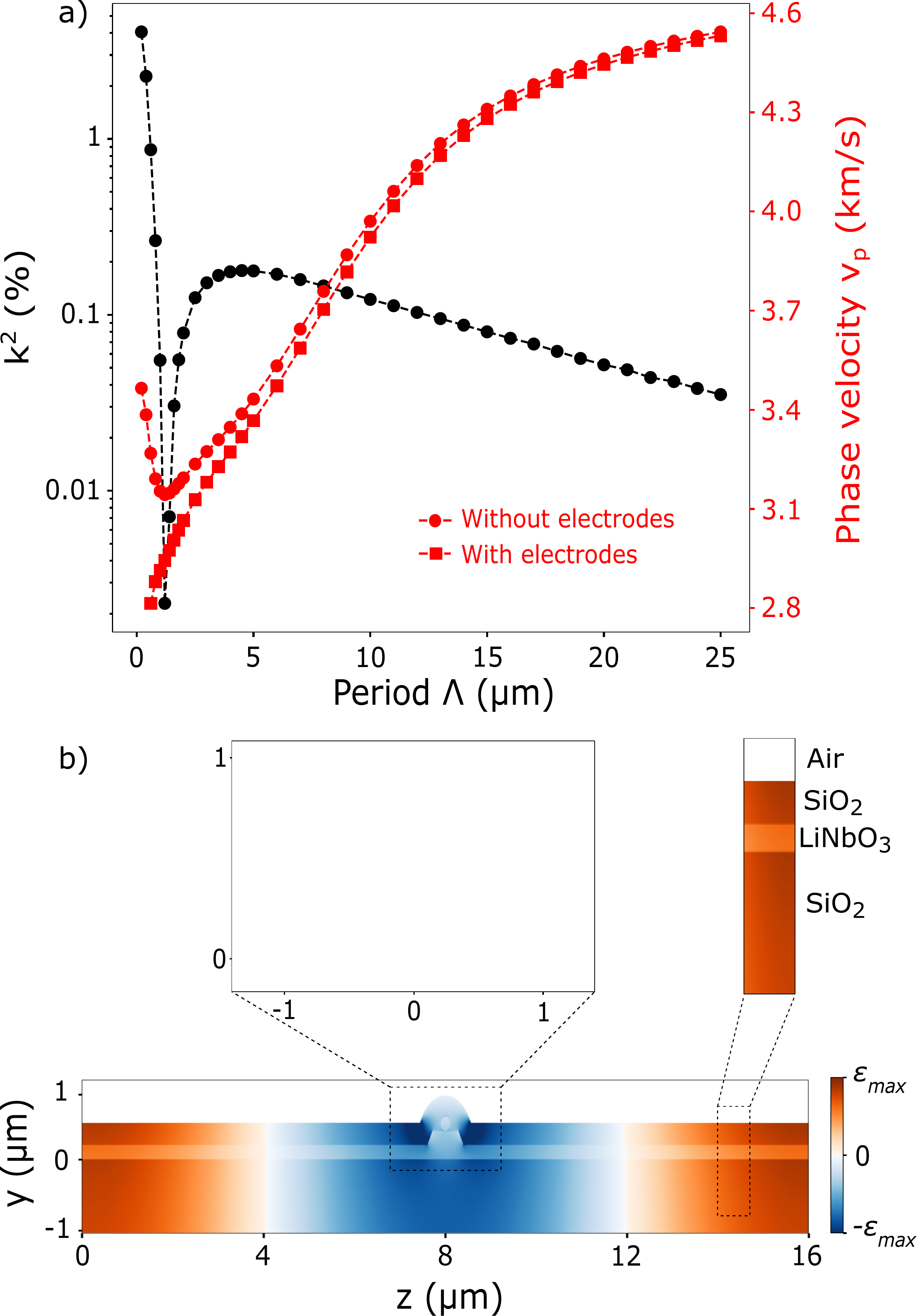}
\caption{\label{fig:k2_and_vp} a) Influence of the IDT period $\Lambda$ on the electromechanical coefficient $k^{2}$ (black) and the Rayleigh wave phase velocity $v_{p}$ (red) for Y-Z LNOI (200 nm \ce{LiNbO_3} - $\SI{3}{\micro m}$ burried \ce{SiO_2}) without \ce{SiO_2} encapsulation. b) Strain profile generated by a SAW with period \SI{16}{\micro m}. The NW is placed on top of a waveguide with the same geometry as in \cref{fig:modes} and is encapsulated by \SI{320}{nm} of \ce{SiO_2}. The simulation axes correspond to that of the crystal.}
\end{figure}

The properties of the surface acoustic wave depend on their propagation medium. For the LNOI under consideration, the Rayleigh wave phase velocity $v_{p}$ and the electromechanical coupling coefficient $k^{2}$ now depend on the thicknesses of the \ce{LiNbO_3} and burried \ce{SiO_2} layers. These two parameters were assessed by 2D FEM simulations with COMSOL in eigenmode (the encapsulating \ce{SiO_2} layer is not considered in this simulation since it is only present in the waveguide region). 
Two electrodes (60 nm of chromium) are placed on the surface of the modeled \ce{LiNbO_3}/\ce{SiO_2}/Si layer stack. By grounding one electrode while setting the other to a floating potential, two propagating modes, one symmetric and one anti-symmetric, are identified. The Rayleigh wave phase velocity is derived as $v_{p} = (f_{S} + f_{A})\cdot\Lambda/2$, where $f_{S}$ and $f_{A}$ are the frequencies of the symmetric and anti-symmetric modes, respectively. The electromechanical coupling coefficient is computed as $k^{2} = 2(v_{p,o}-v_{p,s})/v_{p,o}$, where $v_{p,o}$ and $v_{p,s}$ are the phase velocities when the surface is open and shorted, respectively \cite{Kaletta2014}. 
For $\SI{200}{nm}$ \ce{LiNbO_3} and $\SI{3}{\micro m}$ buried \ce{SiO_2}, the influence of the IDT period is shown in \cref{fig:k2_and_vp}(a). For a period comparable to the \ce{LiNbO_3} thickness, the coefficient $k^{2}$ converges to the bulk value around $\SI{4.1}{\percent}$. Similar optomechanical coupling as demonstrated above is therefore expected. The phase velocity converges to the phase velocity of bulk \ce{LiNbO_3} around $\SI{3490}{m s^{-1}}$ when no physical electrodes are on the surface. Adding the electrodes slows down the wave drastically due to their weight, and the modes are distorted for very short periods. 
Up to $\Lambda < \SI{1.5}{\micro m}$, the wave propagates in the \ce{LiNbO_3} film but partially leaks into the \ce{SiO_2} layer, resulting in a drastic decline of the coefficient $k^{2}$ and slower wave velocities. For $\Lambda > \SI{1.5}{\micro m}$, the phase velocity increases as the wave penetrates deeper into the \ce{SiO_2} to reach the Si substrate which is faster medium than \ce{SiO_2} and \ce{LiNbO_3}. The coefficient $k^{2}$ also increases to a maximum value of 0.17\% due to the waveguiding effect in the \ce{LiNbO_3} and \ce{SiO2} layers, which couples more vibration in these layers \cite{Ma2022}. For $\Lambda > \SI{5}{\micro m}$, this confinement effect declines as the wave propagates deeper in the the silicon, resulting in a $k^{2}$ drop. For a period of \SI{16}{\micro m}, $v_{p}$ = $\SI{4350}{m s^{-1}}$ and $k^{2} = \SI{0.074}{\percent}$. This value of $k^{2}$, although smaller than the bulk scenario, is comparable to that of bulk GaAs ($\approx$ $\SI{0.07}{\percent}$) on which modulation of In(Ga)As QDs has been demonstrated \cite{Weiß2018}. Then, we evaluate the impact of the waveguide and encapsulation oxide ($\SI{320}{nm}$) on the coupling between the SAW and the NW by computing the strain at the center of the NW (\cref{fig:k2_and_vp}(b)). 
Its amplitude drops by $\SI{43}{\percent}$ when the NW is placed on top of the waveguide compared to the case when it is on the surface of a $\SI{200}{nm}$ thick LNOI (Supporting Information S7). This loss mostly originates from the dome of \ce{SiO_2} around the waveguide which creates a local strain minimum. 
Furthermore, we estimate from the simulations that the strain at the center of the encapsulated NW on top of the LNOI waveguide will be 24 fold smaller that the strain in an encapsulated nanowire placed on the surface of the bulk \ce{LiNbO_3}. This $\SI{28}{dB}$ loss translates to a spectral shift of $\SI{0.12}{nm}$ at $\SI{19}{dBm}$ RF power according to \cref{fig:CW_measurement_woSIO2}. 
Therefore, more RF power will be necessary to shift the emission of the NW on LNOI waveguide to partially compensate for this loss. This has the detrimental effect of increasing the heat load on the sample, which can nonetheless be mitigated by RF pulsed excitation (see Supporting Information S2 for further discussion about operation at high RF power). Besides increasing the RF power, the modulation efficiency can be improved in several ways. First, the SAW frequency can be increased, as it allows to reach a larger electromechanical coupling - at $\Lambda = \SI{5}{\micro m}$, $k^{2} = \SI{0.18}{\percent}$, more than twice the value at $\SI{16}{\micro m}$ -, and generates a stronger strain since strain scales linearly with the SAW frequency. Second, the geometry of the IDT can be optimized to launch the SAW only in the direction of the QD instead of the bidirectional emission of the design considered in this study \cite{Ekstrom2017}. Thus, a loss of 3 dB can ideally be avoided. Another option would be to use a focusing IDT and place the QD at the acoustic waist in order to strengthen the acousto-optic interaction \cite{Imany2022}. Third, the local strain minimum in the \ce{SiO_2} dome can be mitigated with a thicker encapsulation layer (see Supporting Information S7).

Other strain-tuning techniques have been reported to shift the QD emission in integrated photonic circuits. Elshaari et al. \cite{Elshaari2018} showed a $\SI{0.8}{nm}$ spectral shift of a nanowire QD emission integrated in a photonic circuit fabricated on top of a PMN-PT substrate. Large voltages ($\SI{600}{V}$) had to be applied to observe this shift, but the power consumption remained low. The strain was equally applied to the source and all circuitry elements, hence limiting selective tuning of individual components. This issue could be overcome by working with thin PMN-PT films \cite{Chen2016}, but their integration with photonic circuitry will likely involve a sophisticated fabrication process and, to the best of our knowledge, has yet to be demonstrated. Capacitive micromechanical systems \cite{Errando-Herranz2020} have also been used to tune the emission of QDs in a III/V platform \cite{Petruzzella2018} and color centers \cite{Meesala2018, Wan2020}. Reasonably low voltages had to be applied to observe a spectral shift, while the power consumption also remained low since it mainly resulted from leakage currents. Nevertheless, this approach relies on fragile suspended membranes and can suffer from a substantial hysteresis \cite{Martin-Sanchez2017}. The tuning scheme presented here will comparatively require a higher power consumption but IDTs can easily be fabricated with high success rate near each emitter to tune them independently.

\section{Conclusion}
In conclusion, we demonstrated the dynamic modulation of the emission of a quantum dot nanowire, transferred to a \ce{LiNbO_{3}} substrate, using surface acoustic waves. The strong optomechanical coupling results in a wide optical tuning range at moderate RF powers, thereby avoiding heating of the sample, charge transport or Stark effect modulation. The purity of the single photon source was preserved during the dynamic modulation process. Adding a \ce{SiO_{2}} encapsulation layer on top of the nanowire does not significantly deteriorate the modulation and presents two benefits. It firmly anchors the nanowire to the substrate and acts as a cladding layer when the nanowire is integrated on a waveguide. Towards this idea, we proposed an architecture to adiabatically transfer the light emitted by the quantum dot into a rib waveguide made from LNOI compatible with strain modulation. The use of a \ce{LiNbO_{3}} thin film combined with an encapsulated waveguide reduces the acousto-optical coupling to a level of performance comparable to that of bulk GaAs for a SAW period of $\SI{16}{\micro m}$. Operating at higher SAW frequency, a better IDT design and a reasonably thicker \ce{SiO_{2}} encapsulation layer are identified as solutions to improve the modulation efficiency. The scheme presented can be extended to other types of nanowire quantum dots, such as GaAs QD in AlGaAs nanowire emitting close to $\SI{780}{nm}$ which can be strain-tuned to match the D2 cycling transition of $^{87}$Rb \cite{Leandro2020}. More generally, this approach is compatible with other emitters like semiconductor quantum dots in tapered nanobeam, or defects in 2D materials and crystals. In addition to the source modulation, the properties of other photonic components such as Mach-Zehnder interferometers and ring resonators can also be acoustically modulated to engineer reconfigurable quantum photonic circuits on LNOI.

\begin{acknowledgments}
The work was partially supported by the Knut and Alice Wallenberg (KAW) Foundation through the Wallenberg Centre for Quantum Technology (WACQT). The authors also acknowledge the support from the European Union’s Horizon 2020 Research and Innovation Programme through the project aCryComm, FET Open Grant Agreement no. 899558.
\end{acknowledgments}


\bibliography{references.bib}

\onecolumngrid
\clearpage
\begin{center}
\textbf{\large Supporting Information}
\end{center}
\setcounter{equation}{0}
\setcounter{figure}{0}
\setcounter{table}{0}
\setcounter{page}{1}
\makeatletter
\renewcommand{\theequation}{S\arabic{equation}}
\renewcommand{\thefigure}{S\arabic{figure}}

\thispagestyle{empty}
\vspace{0.5cm}


\textbf{S1. Power-resolved PL}\\

\begin{figure}[!htbp]
\includegraphics[width=0.45\linewidth]{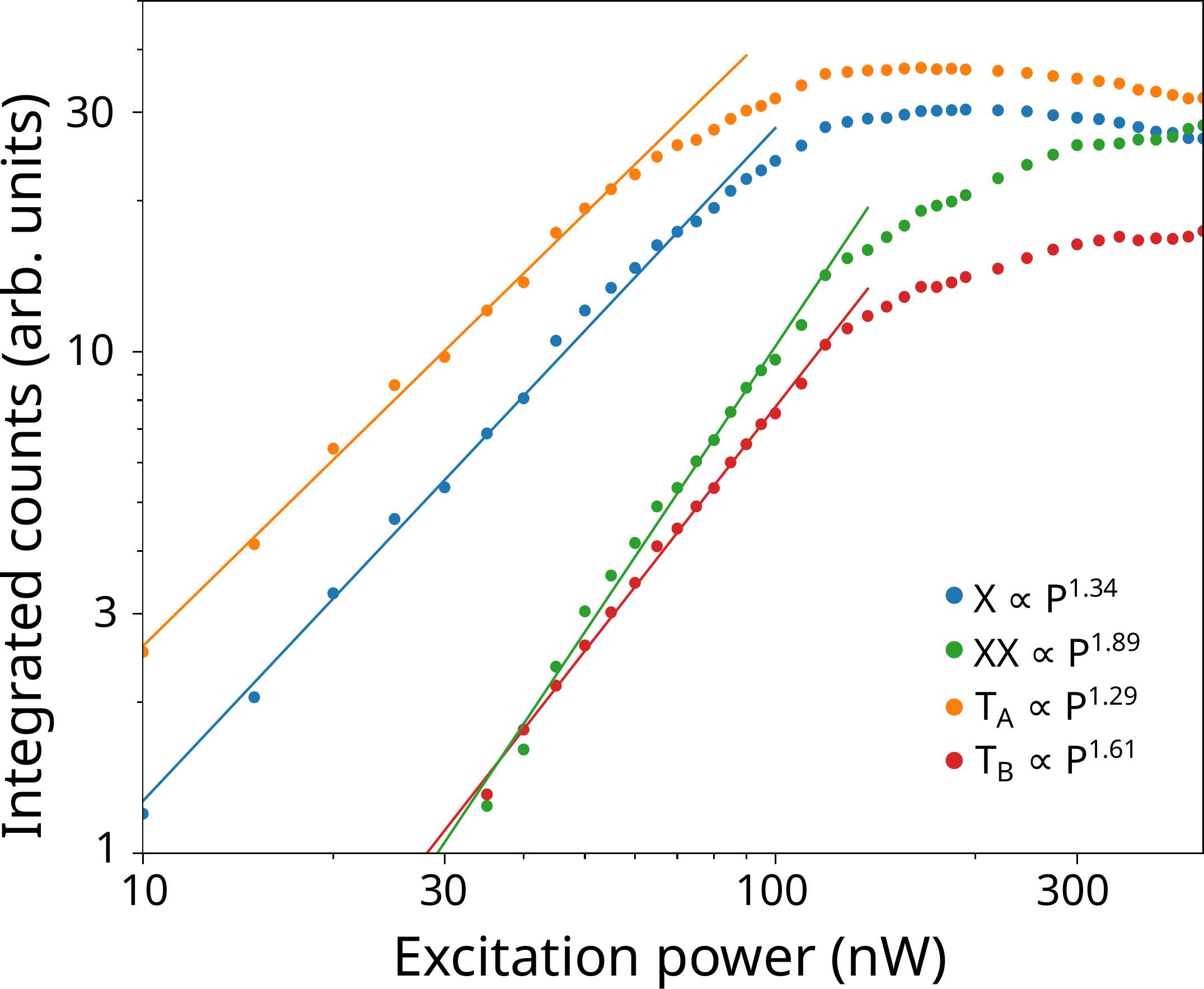}
\caption{\label{fig:power dependence} Power dependent measurement of the 4 emission lines identified in the nanowire quantum dot excited with a HeNe laser. The points are experimental values and the solid lines linear fits.}
\end{figure}

In \cref{fig:power dependence}, the integrated count is shown as a function of the laser excitation power. The blue, green, orange and red datasets correspond to the lines labeled X, XX, TA, TB in the main text, respectively. X and TA exhibit the lowest slopes while XX and TB have the greatest ones [1].\\

\textbf{S2. Spectral modulation at strong microwave power}\\

\begin{figure}[!htbp]
\includegraphics[width=0.5\linewidth]{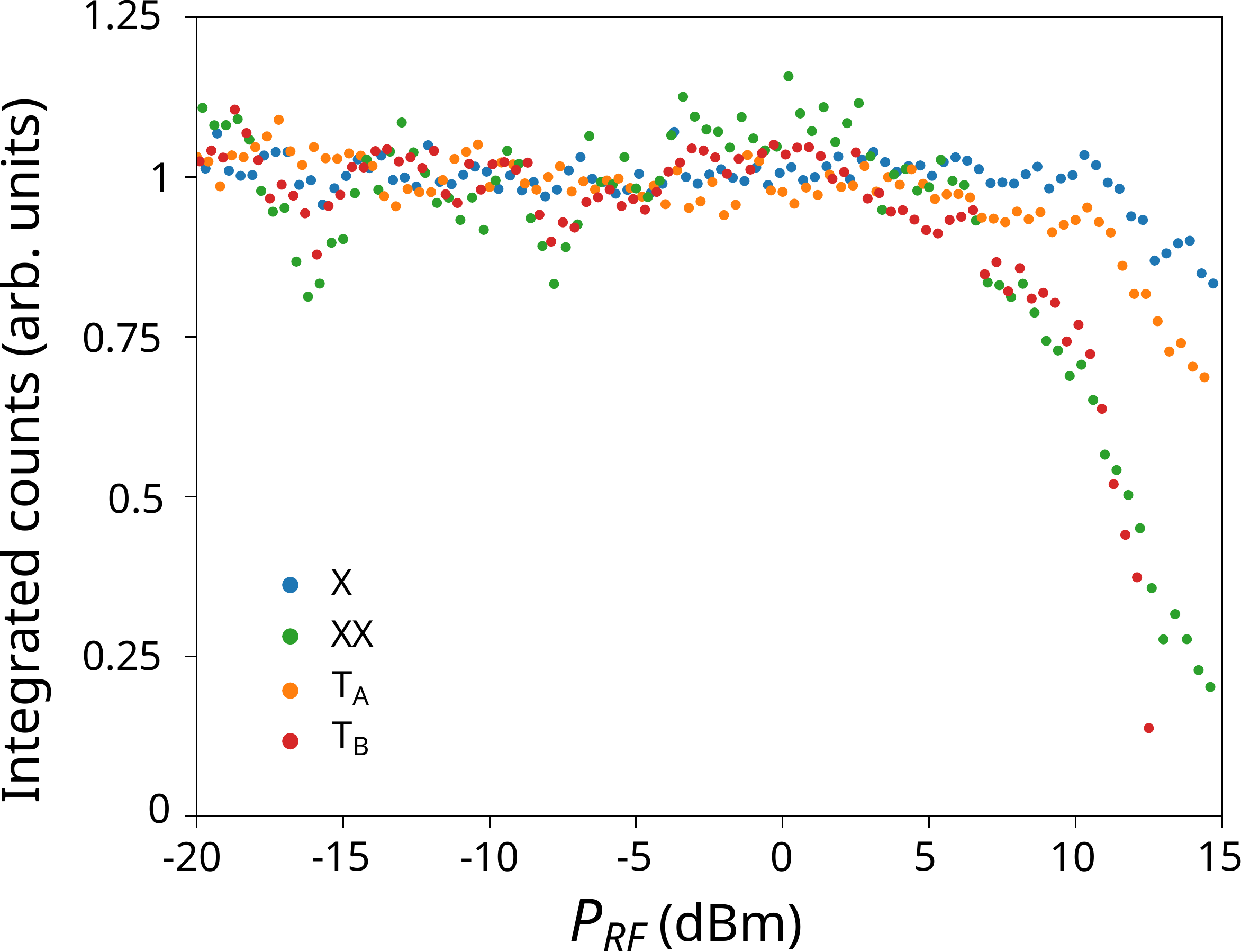}
\caption{\label{fig:integrated counts} Integrated count rates of the four peaks represented in Fig. 1(e) in the main text. Each peak was fitted by a time-integrated oscillating Lorentzian. The intensity starts to drop for all the peaks slightly above PRF = 5 dBm. }
\end{figure}

The trend of Fig.1e) in the main text does not show saturation or a decrease of the tuning amplitude range at the maximum RF power that we applied, which indicates that the spectral shift can be further increased with larger strain fields. However, there is an upper bound to the RF power that can be applied. If we only consider the behavior of the IDT, degradation of the electrodes occurs at large RF power by acoustomigration of the metal grains [2,3], which shifts the resonance frequency over time. At even larger power, the IDT is completely damaged by electrical breakdown. The power level required to initiate these two effects depends on the metal thickness and electrode geometry, but if we consider that the damage is irreversible around 30 dBm [3], this would mean a maximal shift of 11 nm. However, other factors are probably going to further limit the maximum RF power. First, large RF power is going to heat the QD which will manifest as a redshift of the emission lines [4]. For SAWs applied on GaAs substrate, Buhler et al. estimated that 23 dBm continuous RF power would lead to a sample temperature around 55 K. This will result in a drop of performance regarding the single photon source (visibility of the two-photon interferences). This effect can nonetheless be mitigated with pulsed RF drive. Second, we noticed that the integrated count rate starts to drop above 5 dBm as shown in \cref{fig:integrated counts}. A trade-off between the spectral shift and the count rate may have to be made for some applications.\\

\textbf{S3. Additional Hanbury-Brown Twiss measurement at PRF = 4 dBm}\\

\begin{figure}[!htbp]
\includegraphics[width=0.5\linewidth]{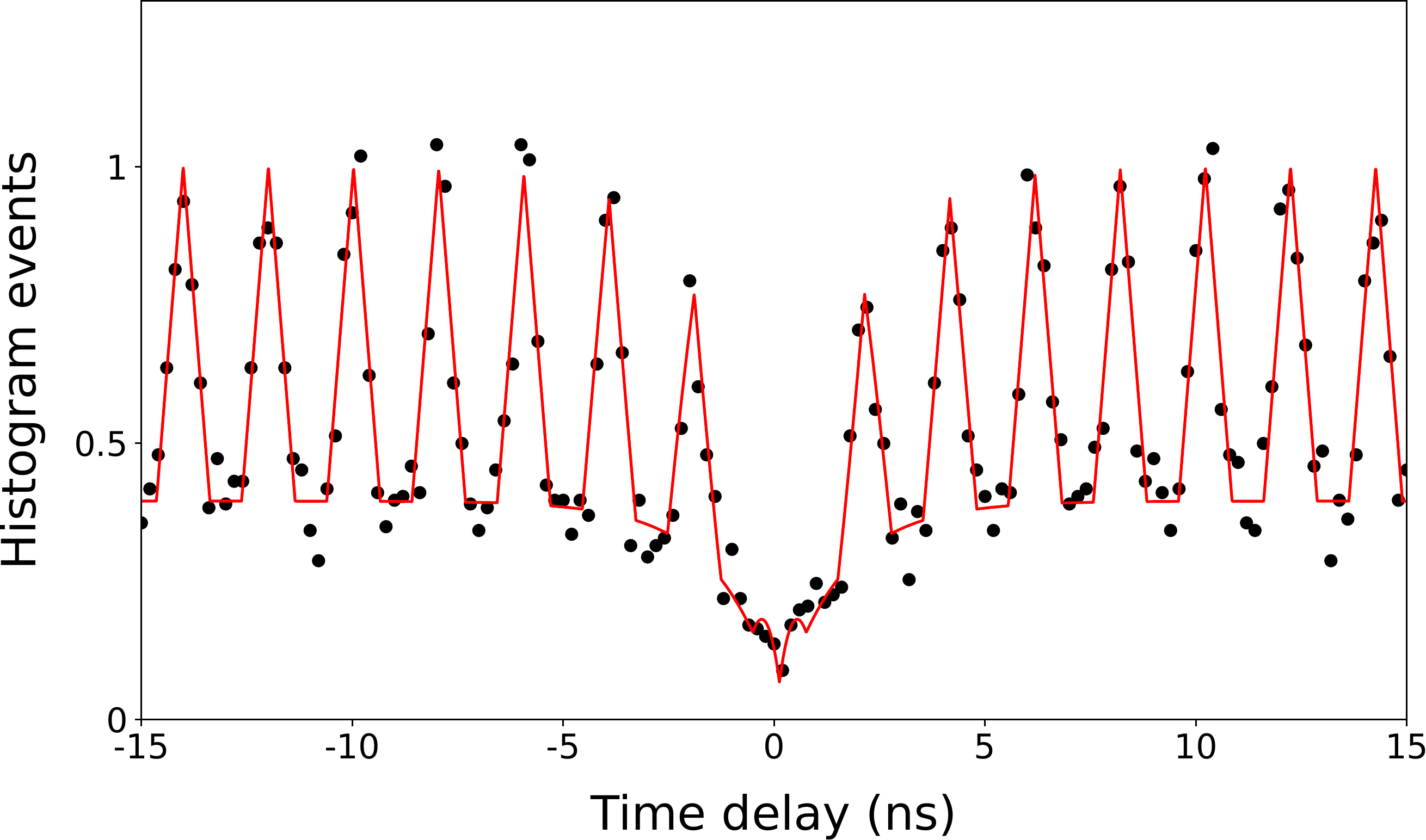}
\caption{\label{fig:g2 center}Hanbury-Brown Twiss experiments of the TA line when the SAW is generated at PRF = 4 dBm and the monochromator set to 846.2 nm. The black dots are experimental values and the red curve is the fit based on the same function as described in the main text.}
\end{figure}

\cref{fig:g2 center} shows the Hanbury-Brown Twiss experiments of the TA line when the acoustic field induced by the SAW is minimum. The period of the pulses is 2.019 ns, which is half of the SAW period since the modulated TA line falls in the band pass of the monochromator twice. The value g$^{2}$(0) = 0.066 is larger than the one shown in the main text, but can be improved with longer integration time.\\

\textbf{S4. Insertion loss before and after \ce{SiO_{2}} deposition}\\

\begin{figure}[!htbp]
\includegraphics[width=0.5\linewidth]{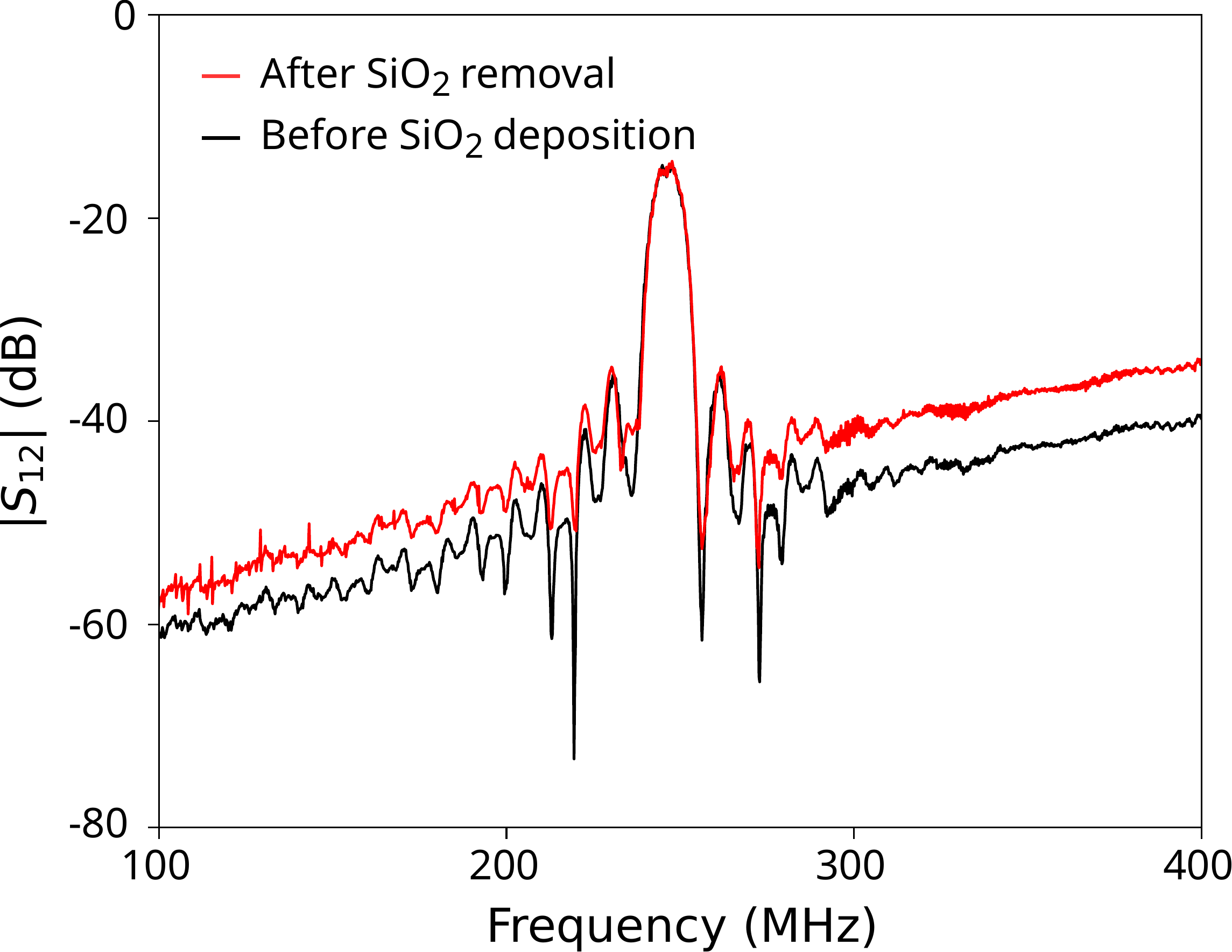}
\caption{\label{fig:scattering parameters} Insertion loss of the delay line before oxide deposition (black) and after oxide removal on the IDT (red). The spectra were acquired when the sample was mounted in the cryostat and cooled to 1.6 K.}
\end{figure}

To ensure that the IDT performance was not altered after removing the \ce{SiO_{2}} covering it, the insertion loss |S21| was compared before and after \ce{SiO_{2}} removal (\cref{fig:scattering parameters}). The resonance peak at 245 MHz has the same linewidth and amplitude, showing that the device performance has not been affected by the process in this region. The background outside the resonance is higher by approximately 5 dB after oxide removal, but this change does not affect the device operation. We note that the resonances found optically within the main lobe due to the acoustic cavity formed by the delay line are surprisingly not present in this electrical measurement.\\

\textbf{S5. Strain profile for a thick encapsulation layer}\\

\begin{figure}[!htbp]
\includegraphics[width=0.8\linewidth]{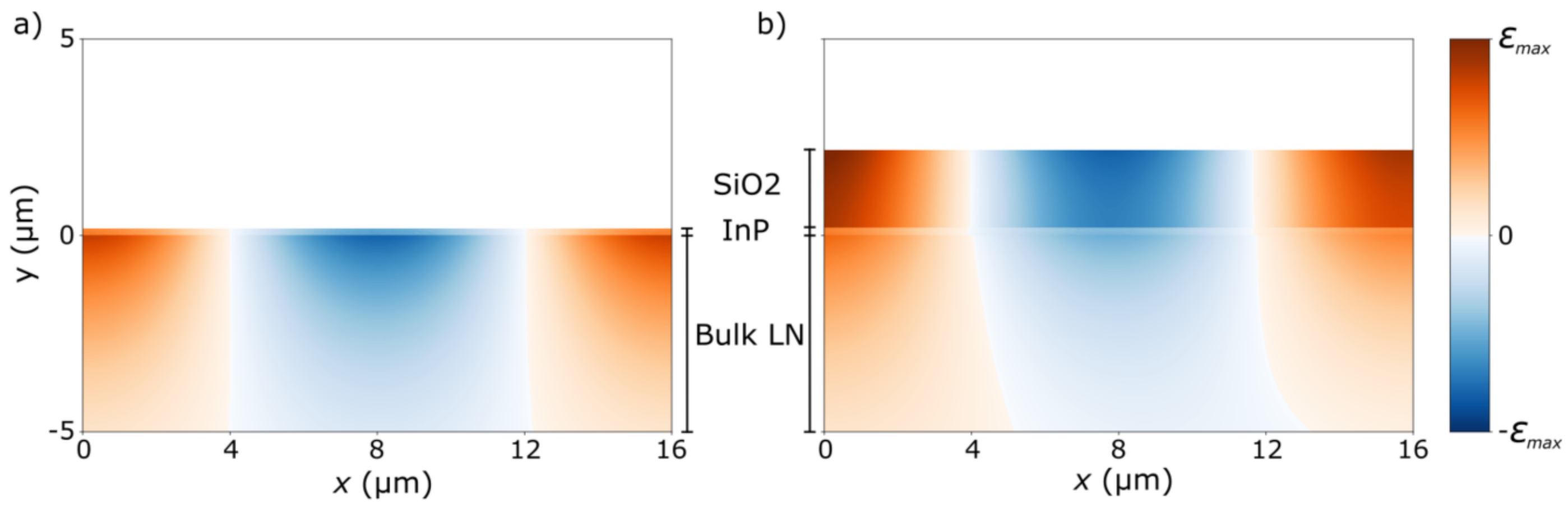}
\caption{\label{fig:strain_field_2um_encapsulation} Strain generated by the SAW (period 16 um) on a bare 128$^{\circ}$ Y-X cut LN substrate/200nm InP stack (a) and with a 2 um \ce{SiO_{2}} encapsulation layer (b).}
\end{figure}

We simulated the displacements generated by the SAW on a 128$^{\circ}$ Y-X cut LN using a finite element method (COMSOL). Then, the strain is computed as $\varepsilon = \frac{\partial u_x}{\partial x} + \frac{\partial u_y}{\partial y}$, where $u_x$ and $u_y$ are the displacements along the x-axis of the crystal and the direction normal to the surface, respectively.
Without encapsulation, the SAW is localized at the surface of the bulk/InP stack (\cref{fig:strain_field_2um_encapsulation}(a)). Adding \ce{SiO_{2}} on the surface disrupts this localization as the wave also propagates in the encapsulation layer. This effect can be well visualized for thick encapsulation layer as shown in \cref{fig:strain_field_2um_encapsulation}(b) for a 2 um deposition. As the wave propagates at the surface of the \ce{SiO_{2}}, the strain in the now burried InP layer decreases.\\

\textbf{S6. Simulations of the nanowire-waveguide optical coupling}\\

\begin{figure}[!htbp]
\includegraphics[width=0.8\linewidth]{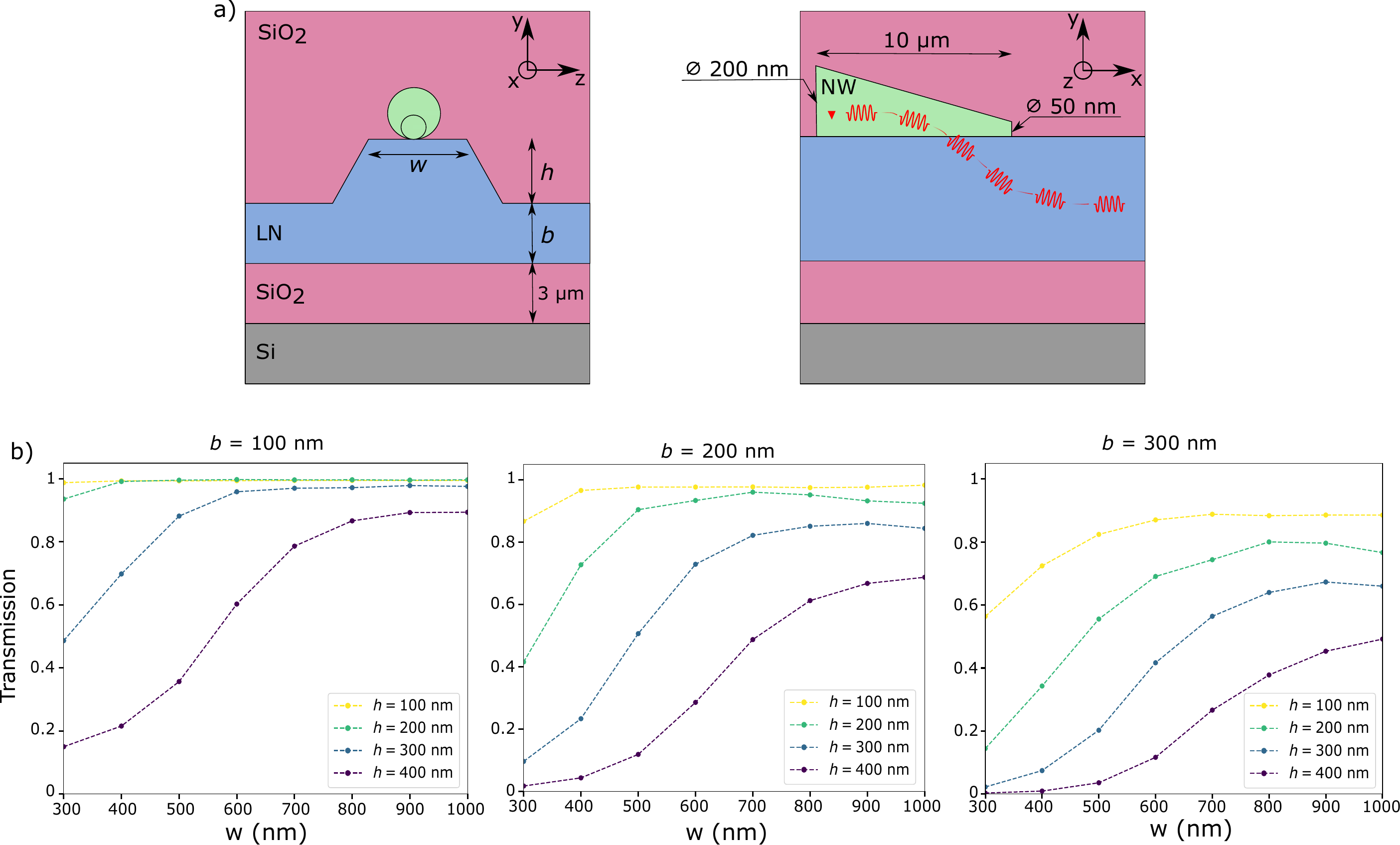}
\caption{\label{fig:sketch LNOI} Parametric study of the LNOI waveguide geometry.  a) YZ view and YX view of the geometry and layer stack. The sidewall angle of ridge is fixed to 60°. b) The transmission, representing the coupling efficiency between the nanowire and the waveguide modes, is computed at 850 nm as a function of the waveguide width w, waveguide height h and base thickness b.}
\end{figure}

The coupling efficiency between the nanowire quantum dot and the LNOI waveguide was simulated with an eigenmode expansion solver (Lumerical). The geometry of the model is shown in \cref{fig:sketch LNOI}(a) and the LNOI thickness, waveguide height and width were varied. The input port (port 1) was set to the fundamental TE mode of the nanowire and the output port (port 2) was the fundamental TE mode of the LN waveguide. All the materials were considered loss-less. Better coupling is obtained for wider ridge with a thinner base and a smaller height (\cref{fig:sketch LNOI}(b)).\\

\textbf{S7. Additional strain profiles for the LNOI architecture}\\

\begin{figure}[!htbp]
\includegraphics[width=0.7\linewidth]{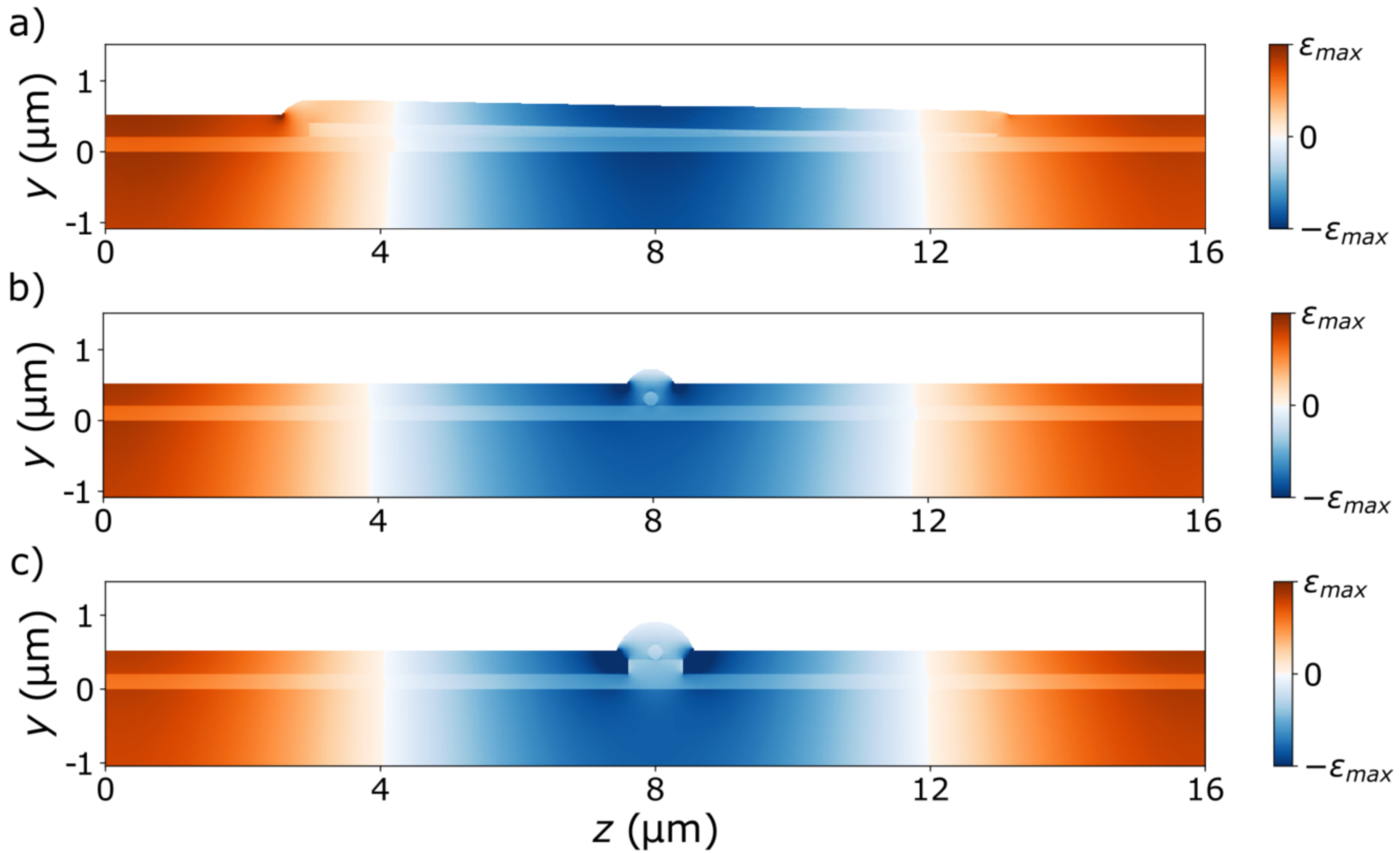}
\caption{\label{fig:strain_field_horizontal-perpendicular} Strain induced by the SAW launched along the z-axis of the LNOI cladded with 320 nm of \ce{SiO_{2}} and driven at 271.8 MHz (period 16 um) in three different configurations. a) The nanowire QD is on the 200 nm LNOI and parallel to the z-axis. b) The nanowire QD is on the 200 nm LNOI and perpendicular to the z-axis. c) The nanowire QD is on a LNOI waveguide with straight sidewalls and perpendicular to the z-axis. The scale bar of the strain is the same for all three subfigures. The axes correspond to those of the Y-cut LNOI.}
\end{figure}

The strain is computed as $\varepsilon = \frac{\partial u_y}{\partial y} + \frac{\partial u_z}{\partial z}$, where $u_y$ and $u_z$ are the displacements along the y-axis  and z-axis of the Y-cut LN crystal, respectively.
The influence of the orientation of the nanowire QD with respect to the SAW propagation direction was investigated with the two extreme cases, parallel and perpendicular. We considered the same Y-cut LNOI as described in the main text as well as the same nanowire geometry as shown in \cref{fig:sketch LNOI}(a), with a 320 nm \ce{SiO_{2}} encapsulation (b = 200 nm, h = 200 nm, w = 800 nm). For both orientations, the strain field generated by the SAW couples to the nanowire QD when it is placed on the surface of a 200 nm thick LNOI (\cref{fig:strain_field_horizontal-perpendicular}(a) and (b)). The strain at the center of the nanowire in the perpendicular case is 10 \% larger than in the parallel case. Therefore, the nanowire orientation does not seem to have a significant impact on the final acousto-optical modulation. 
To quantify the influence of the sidewall angle of the ridge waveguide on the strain field, we considered the worst-case scenario where the waveguide has straight walls (\cref{fig:strain_field_horizontal-perpendicular}(c)). The strain field amplitude at the center of the nanowire is 98 \% to that of the case with 60$^{\circ}$ sidewall angle.\\

\begin{figure}[!htbp]
\includegraphics[width=0.7\linewidth]{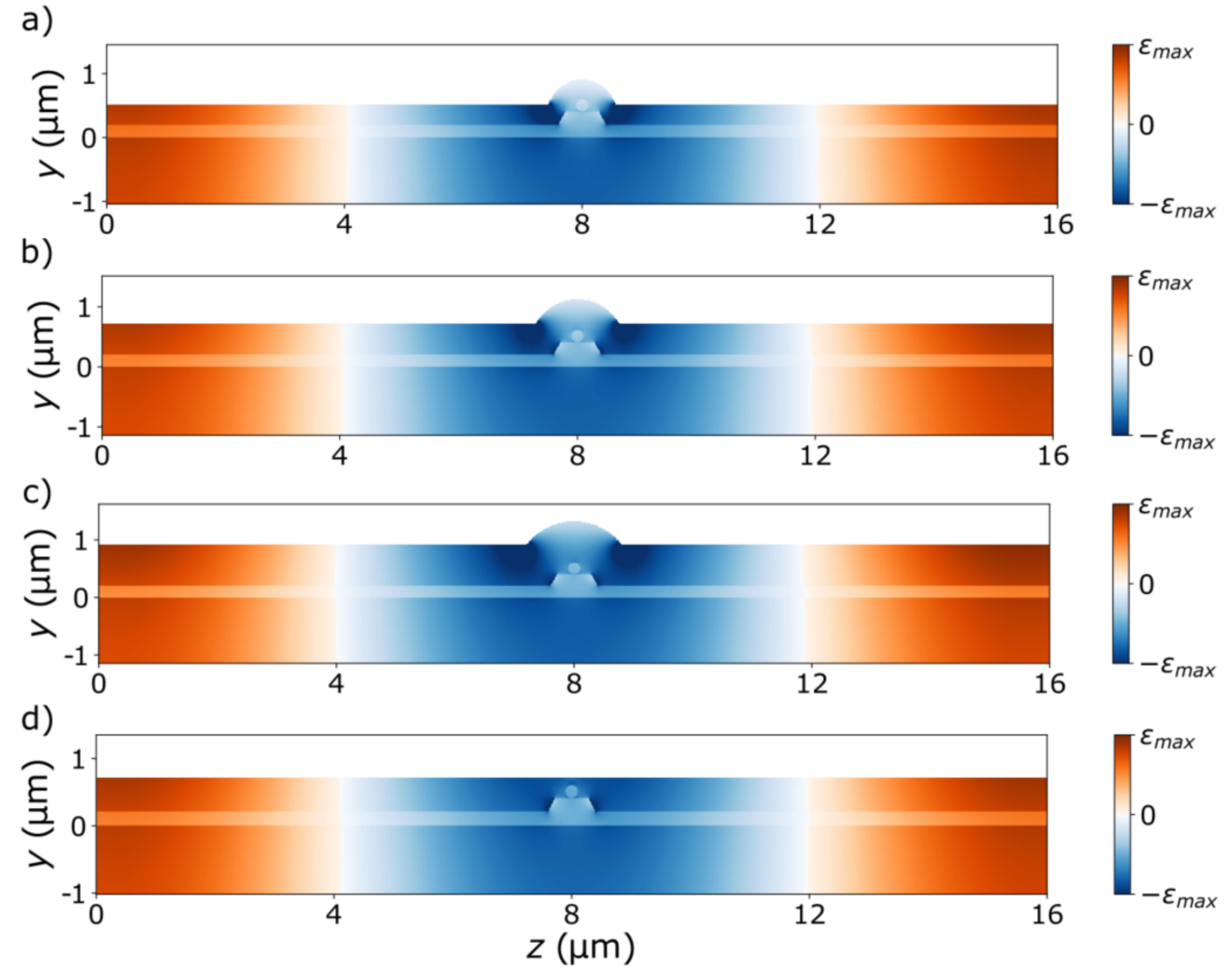}
\caption{\label{fig:strain_field_dome_height} Strain induced by the SAW (period 16 um) for 320 nm (a), 520 nm (b) and 720 (nm) \ce{SiO_{2}} encapsulating layer. The crystal axes correspond to those of the Y-cut LNOI. In d), the dome resulting from the conformal deposition has not been modelled as an extreme case. The scale bar of the strain is the same for all four subfigures.}
\end{figure}

As discussed in the main text, the encapsulation dome of \ce{SiO_{2}} on top of the waveguide (\cref{fig:strain_field_dome_height}(b)) due to the conformal deposition decreases the strain inside the NW. Thicker encapsulation layers can alleviate this effect because the region of local minimum strain in the dome is further above the nanowire. It is mentioned in the main text that, for a 320 nm encapsulation layer, the loss of strain at the center of the nanowire on top of the waveguide compared to the case where the nanowire is on the bare 200 nm thick LNOI is 43\%. This loss decreases to 28\% for 520 nm (Fig S8(b)) and 17\% for 720 nm (\cref{fig:strain_field_dome_height}(c)) encapsulation. This thicker encapsulation remains in the range where the leakage of the SAW to the \ce{SiO_{2}} encapsulation is negligible. In the extreme case where the dome is not considered as in \cref{fig:strain_field_dome_height}(d), the loss of strain in the nanowire becomes minor.

\textbf{S8. Linewidth of the peak TA before and after \ce{SiO_{2}} deposition}\\

\begin{figure}[!htbp]
\includegraphics[width=0.45\linewidth]{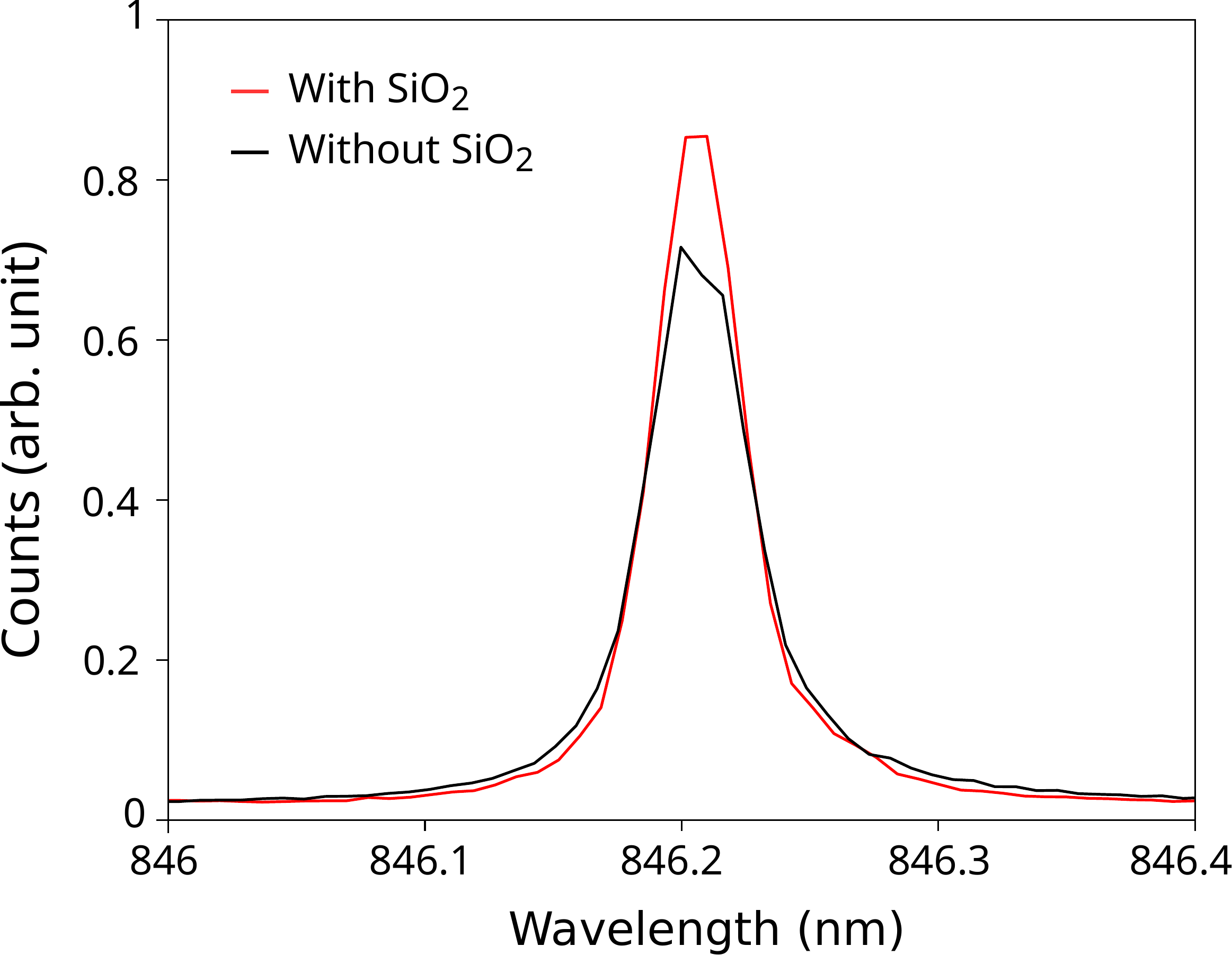}
\caption{\label{fig:linewidths} Spectral linewidth of the TA line before (black) and after (red) oxide deposition. The red curve has been shifted by 1.92 nm to compensate for the constant blueshift introduced by the oxide. No significant broadening could be observed. The quantum dot was excited continuously with a HeNe laser at 150 nW.}
\end{figure}

The linewidth measured on the spectrometer did not show noticeable change after \ce{SiO_{2}} encapsulation, as shown in \cref{fig:linewidths}.\\

\textbf{S9. Lifetime measurement of TA}\\

\begin{figure}[!htbp]
\includegraphics[width=0.5\linewidth]{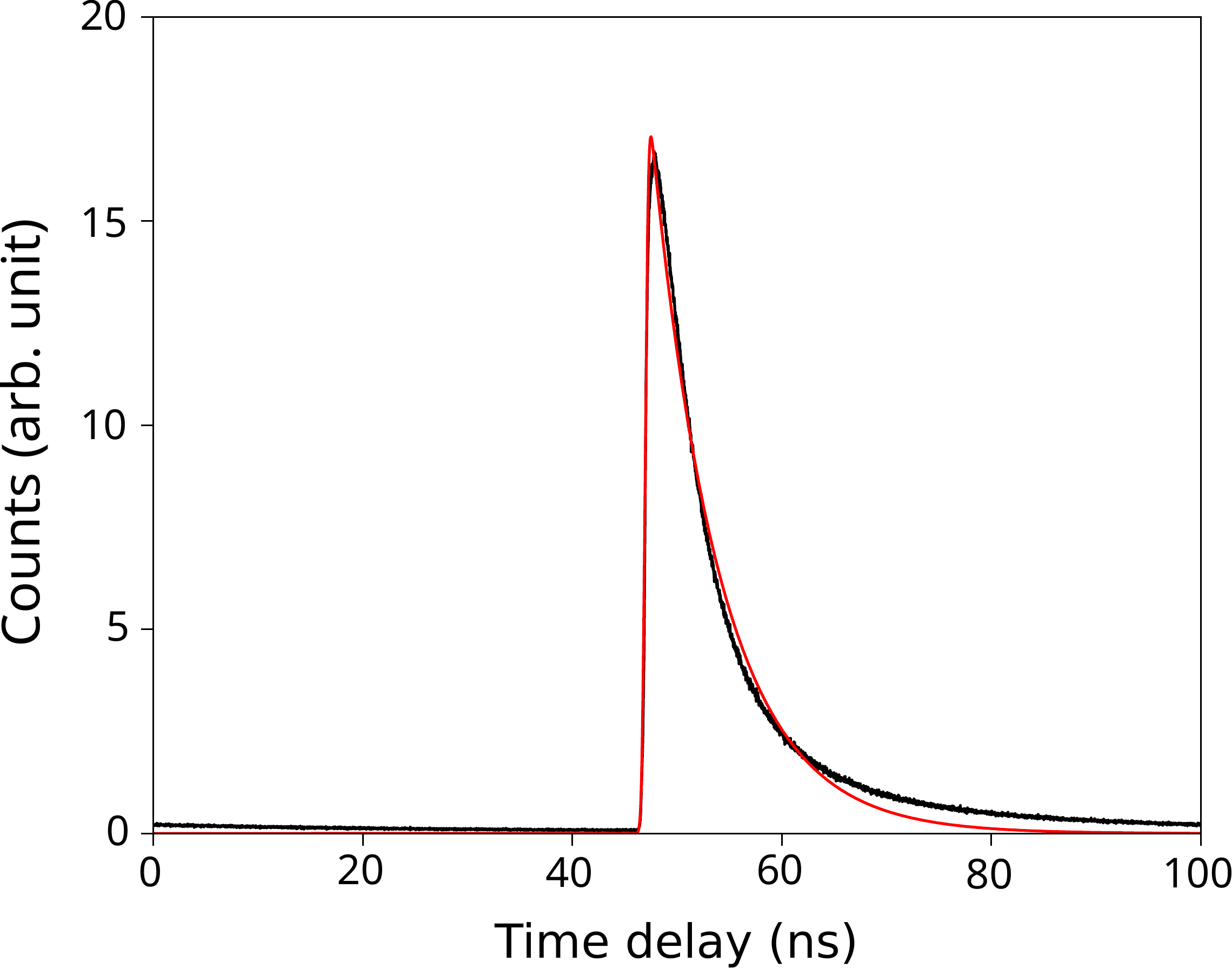}
\caption{\label{fig:lifetime} Lifetime measurement (black line) of the TA line and fitting function (red).}
\end{figure}

The lifetime of the TA line was measured with pulsed HeNe laser at 5 MHz at 150 nW. \cref{fig:lifetime} shows the experimental data together with a fit consisting of an exponential convoluted with the instrument response function (gaussian with a standard deviation of 244 ps). The extracted lifetime is 6.52 ± 0.1 ns.\\

\textbf{S10. Strain modulation at 735 MHz}\\

\begin{figure}[!htbp]
\includegraphics[width=0.9\linewidth]{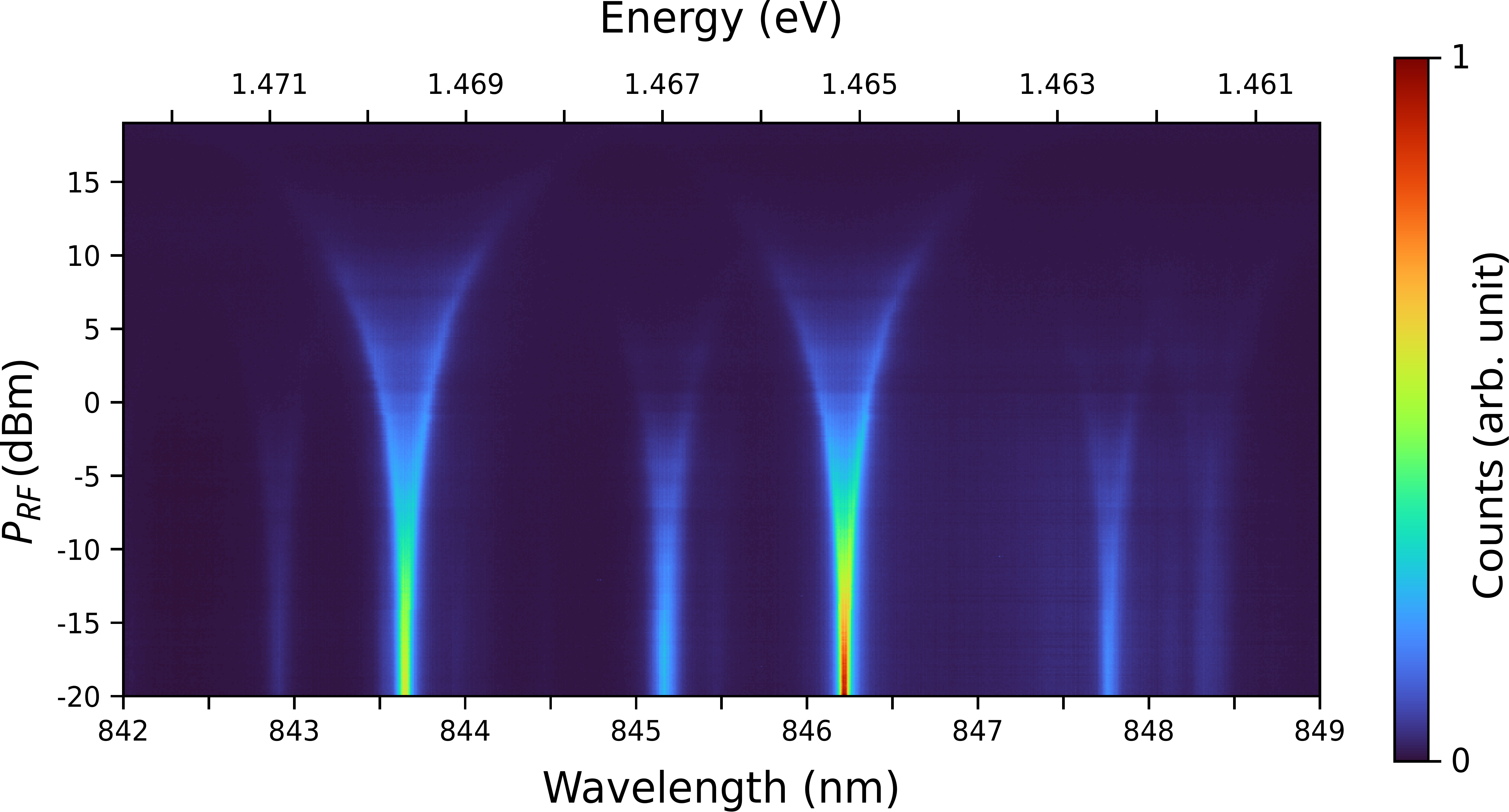}
\caption{\label{fig:H3 power sweep} Strain-induced energy broadening of the emission lines of the nanowire quantum dot by driving the IDT at the third harmonic 735 MHz. The nanowire QD shows a response similar to the modulation with the fundamental mode. The quantum dot was excited with a HeNe laser at 150 nW.}
\end{figure}

The spectral shift induced by the SAW generated at the third harmonic 735 MHz of the IDT is shown in \cref{fig:H3 power sweep}.\\

\textbf{S11. Estimation of the maximum number of single emitters tuned with SAW on the same chip}\\

In this study, the emitter was placed in a delay line formed by two IDTs. Ultimately, only one IDT per emitter is enough to drive the SAW for independent tuning. If we consider the Y-cut, the IDTs should be placed next to one another along the X direction to generate SAWs along the Z direction. Two IDTs should not face each other in the Z direction to avoid overlap of the two generated SAWs. If the same acoustic aperture as in the text (180 um) is kept and each IDT is separated from its neighbors by 70 microns, 4 emitters can be integrated per mm. By reducing the acoustic aperture to diminish the footprint of the IDT while preserving its performance, the number of emitters could be pushed to 5 per mm.\\

\textbf{References}\\

[1]	Weiß, M.; Schülein, F. J. R.; Kinzel, J. B.; Heigl, M.; Rudolph, D.; Bichler, M.; Abstreiter, G.; Finley, J. J.; Wixforth, A.; Koblmüller, G.; Krenner, H. J. Radio Frequency Occupancy State Control of a Single Nanowire Quantum Dot. Journal of Physics D: Applied Physics, 2014, 47, 394011.

[2]	Schmidt, H.; Kunze, R.; Weihnacht, M.; Menzel, S. Investigation of acoustomigra-
tion effects in Al-based metallizations. Proceedings of the IEEE Ultrasonics Symposium
2002, 1, 415–418.

[3]	Paquit, M.; Djoumi, L.; Vanotti, M.; Soumann, V.; Martin, G.; Blondeau-Patissier, V.;
Baron, T. Displacement of Microparticles on Surface Acoustic Wave Delay Line Using
High RF Power. IEEE International Ultrasonics Symposium, IUS 2018, 2018-Octob.

[4]	Bühler, D. D.; Weiß, M.; Crespo-Poveda, A.; Nysten, E. D.; Finley, J. J.; Müller, K.;
Santos, P. V.; de Lima, M. M.; Krenner, H. J. On-chip generation and dynamic piezo-
optomechanical rotation of single photons. Nature Communications 2022, 13, 1–11.


\end{document}